\documentstyle[prl,aps,preprint]{revtex}

\begin{document}

\preprint{TAUP-2344-96, LPTENS 96/11, ULB-TH-96/08, hep-th/9606106}
\title{Black Hole Horizon Fluctuations}
\author{A.Casher $^ a$\thanks{Email address: ronyc@post.tau.ac.il}, 
F. Englert $^a$$^b$\thanks{Email address: fenglert@ulb.ac.be}, 
N.Itzhaki $^a$\thanks{Email address: sanny@post.tau.ac.il},
S.Massar $^a$\thanks{Email address: massar@post.tau.ac.il},
and R. Parentani $^c$\thanks{Email address: parenta@physique.ens.fr}}
\address{$ ^a$ Raymond and Beverly Sackler Faculty of  Exact Sciences,
School of Physics and Astronomy,
\\Tel Aviv University, Ramat Aviv, 69978, Israel
\\
$^b$ Service de Physique Th\'eorique, Universit\'e Libre de Bruxelles,
\\ Campus Plaine, C.P. 225, Bd du Triomphe, B-1050 Brussels, Belgium
\\
$^c$ Laboratoire de Physique Th\'eorique de l'\' Ecole
Normale Sup\'erieure
\\(Unit\'e propre de recherche du C.N.R.S.
associ\'ee \`a l'ENS et \`a l'Universit\'e de Paris Sud.)
\\24 rue Lhomond
75.231 Paris CEDEX 05, France}

\maketitle
\newcommand{\be}{\begin{equation}}
\newcommand{\ee}{\end{equation}}
\newcommand{\leqx}{\,\raisebox{-1.0ex}{$\stackrel{\textstyle <}
{\sim}$}\,}
\newcommand{\geqx}{\,\raisebox{-1.0ex}{$\stackrel{\textstyle >}
{\sim}$}\,}
\newcommand{\th}{\theta}
\newcommand{\va}{\varphi}
\newcommand{\de}{\Delta}
\newcommand{\x}{\tilde{x}}
\newcommand{\bl}{\hspace{-.65cm}}
\newcommand{\p}{\partial}
\newcommand{\w}{Schwarzschild $\:$}
\newcommand{\cb}{centrifugal barrier $\:$}
\newcommand{\s}[1] {\mid \! \! {#1} \rangle}  

\begin  {abstract}
It is generally admitted that gravitational interactions become large  
at an invariant distance of  order $1$ from
the black hole horizon.
We show that 
due to the ``atmosphere'' of high angular
particles near the horizon strong gravitational interactions
already occur at an invariant distance of the order of $\sqrt[3]{M}$.
The implications of these results for the origin of black hole
radiation, 
the meaning of black hole
entropy and the information puzzle are discussed.
\end{abstract}

\newpage
\section{ Introduction}

There are at least three related unsolved problems concerning
quantum black hole physics: 

-the origin of the Hawking radiation \cite{haw1};

-the meaning of the black hole entropy \cite{bek1,bek2,bek3};

-the information loss puzzle \cite{haw}.

All of these issues are connected to the large red shift 
near the horizon which entails the appearance in the free 
field theory of transplanckian 
frequencies\cite{un1}\cite{jac1}\cite{jac2}.

The appearance of transplanckian frequencies  
suggest that even though the Riemann Tensor near the horizon 
is small $R_{hor}\simeq \frac{1}{M^2}\ll 1$
\footnote{In  units where $G=c=\hbar=1$.}, there are strong interactions 
near the horizon \cite{th90}.
As originally stressed by t'Hooft this 
puts into question Hawking's original derivation of black 
hole radiance and suggests the possibility that
the
spectrum of emitted particles could differ 
from the exact thermal spectrum.
This in turn could imply  
that the information is encoded in the correlations between
 Hawking quanta \cite{th90}. 
Other possibilities are that the information
 is either lost \cite{haw} or is encoded in the correlations
 between Hawking radiation and 
a remnant \cite{aha}. Furthermore the  transplanckian
 frequencies 
are also related to the divergence of the field entropy near
 the horizon\cite{th85}.
The strong interactions near the horizon suggest 
that there is 
a dynamical cutoff near the horizon\cite{th94}.
A concrete realization of this idea is  the concept
of the stretched horizon \cite{sus} which is a very hot 
membrane, just outside the event horizon, that can absorb, 
thermalize, and emit information.

Several arguments suggest 
that the strong gravitational 
interactions occur at an invariant distance of the order of $1$
(by invariant distance we mean the distance on constant $t$ surfaces
$\rho = \int_{2M}^r ds \simeq \sqrt{8 M (r-2M)}$):

-1. In various works not directly related to black hole physics, 
it was claimed  that the minimal scale  in quantum gravity is 
$1$\cite{ONE}. If so then it is meaningless
 to describe the region near the horizon with accuracy larger
 then $1$.

-2. The local temperature  is $T_{loc}=\frac{1}{\rho}$, so 
for  $\rho =1$  the temperature is Planckian   for which the 
conventional
 description of physics is probably incorrect.

-3. The mean time between successive emissions of Hawking quanta is $M$ and the
 energy of the emitted particle is $\frac{1}{M}$.
 This  means that as a quantum mechanical  system the black hole
 has a width $\frac{1}{M}$.
The invariant distance between $R=2(M-\frac{1}{M})$ and $R=2M$
 in the gravitational background of a black hole with mass
 $M-\frac{1}{M}$ is $1$.

-4. The number of high angular momentum particles between 
$\rho $ and $r \simeq 3M$
 is $N(\rho)\simeq \frac{M^2}{\rho ^2}$ thus for $\rho =1$ the
 entropy of
 these particles is of the order of the Bekenstein-Hawking
 entropy \cite{th85}.

On the other hand several authors have suggested, using different
approaches, that the
gravitational interactions become strong at invariant distances $\rho$
much larger than 1: York considered quantum fluctuations of the
gravitational normal modes\cite{yor}, Jacobson's arguments where based on
thermodynamic analogies\cite{jac2}, Sorkin considered the vacuum fluctuations
of a scalar field\cite{sor}, and  Englert argued that   
gravitational effects which would tame the
transplanckian fluctuations  had to occur at distance much larger than $\rho=1$:
they  would simultaneously invalidate the conventional mechanism of Hawking
emission and prevent, at least in the reconstructed history available to the
external observer, the collapse of the star\cite{eng}. 
However all of these arguments are based on some
questionable assumptions which have not been widely accepted.

The aim of this article is to address the above debate. Our analysis
is based on the properties of the atmosphere of high angular
momentum particles which surround the horizon.
Let us  recall that this atmosphere arises in the reexpression of the
Unruh vacuum state (the state of the field after 
the radiation has settled into its steady state\cite{Unruh76}) 
as
a thermal density matrix of \w quanta:
\begin{equation}
\vert 0_U \rangle
= \prod_{\omega ,l,m}
\exp -\left ( e^{-8 \pi \omega M} a^\dagger_{\omega , l, m}
b^\dagger_{\omega , l, m}\right) \vert 0_B \rangle \label{Unruhvac}
\end{equation}
where $\vert 0_U \rangle$ is the Unruh vacuum and  $\vert 0_B \rangle$
is the Boulware vacuum.
The operator $a^\dagger_{\omega , l, m}$ creates  
 an outgoing \w quantum of energy $\omega$ and angular momentum $l,m$;
$b^\dagger_{\omega , l, m}$ 
creates the partner of this \w quantum and  
lives beyond the horizon.
Upon tracing over these partners one obtains that 
Unruh vacuum is a thermal density matrix for the \w 
quanta, with temperature $(8\pi M)^{-1}$. 

We present three complementary arguments which show that due to the
presence of this atmosphere gravitational interactions become large
at $\rho \simeq \sqrt[3]{M}$. Therefore at this scale Hawking's assumptions of a free
field propagating on a given classical background breaks down.
In particular  the decomposition Eq. (\ref{Unruhvac}) which results from these
 hypothesis will no longer be valid for $\rho < \sqrt[3]{M}$.

Our first argument is thermodynamic in character and relies
 only on the thermal
energy fluctuations in the atmosphere and on the gravitational
analogue of Gauss's law. The second argument 
is based on the gravitational interactions between an incoming
particle and the atmosphere, and shows that these interactions
 can no
longer be neglected for $\rho < \sqrt[3]{M}$. 
The third argument 
is concerned with the unitarity problem. We show that the
information carried by an incoming particle gets encoded in
the state of the atmosphere at $\rho \simeq \sqrt[3]{M}$.
These arguments all lead to the same
minimal distance $\rho =\sqrt[3]{M}$.

 The appearance of strong fluctuations at scales
 much larger
than $\rho = 1$ has important implications for the origin of the
Hawking radiation, the interpretation of the black hole 
entropy and
the unitary problem. These aspects are discussed in the last 
section of the paper.

\section{Horizon fluctuations}
We first review the properties of the high angular momentum particles
which make up the atmosphere.
To this end we recall
the wave equation for a scalar field in the Schwarzschild metric:
\be \left(\frac{\p ^2}{\p t^2}-\frac{\p ^2}{\p r_*^2}\right)
\phi +\left(1-\frac{2M}{r}\right) \left(\frac{2M}{r^3}+
\frac{l(l+1)}{r^2}\right)\phi=0.
\label{KG}\ee
where $r^* = r + 2M \ln (\frac{ r- 2M}{2M})$.
The \cb  
is attractive for $r<3M$.
This means that particles with high angular momentum can be
 trapped in the region between the horizon and $r<3M$. 
The   tunneling through the angular momentum barrier may be
 neglected for   all but the lowest angular momentum modes \cite{can}.
 From Eq.(\ref{KG}) we see that a particle can reach a radius 
$r$ only if 
\be \omega^2 >\left(1-\frac{2M}{r}\right)\left(\frac{2M}{r^3}+
\frac{l(l+1)}{r^2}\right)\label{maxr}\ee
where $\omega$ is the Schwarzschild energy, i.e., the eigenvalue of
$i\partial_t$.
For Hawking radiation the typical energy is $\omega \simeq\frac{1}{M}$.
 This implies
\be l^2 \leqx\frac{M^2}{\rho ^2},\label{maxl}\ee
where $\rho$ is the invariant distance from the horizon in 
Schwarzschild coordinate, namely,
\be  \rho =\int_{2M}^{R} ds=\int_{2M}^{R}\frac{dr}
{\sqrt{1-\frac{2M}{r}}}\simeq\sqrt{8M(R-2M)}\label{rho}\ee
Since the degeneracy for each $l$ is $2l+1$, the number of 
modes that can reach   $\rho $  is 
\be N(\rho)\simeq\frac{M^2}{\rho ^2}.\label{Nrho}\ee

As mentioned in the introduction, 
all of these particles are in a thermal distribution at 
the global Hawking temperature $1/8\pi M$.
The average number of particles in the thermal atmosphere  
is estimated by noting that 
the emission rate for
each mode is $1/M$ and the time it takes for a photon which passes
$\rho$ to fall back to $\rho $  after being reflected by the \cb is 
$O(M \ln M)$. Thus the average number of particles 
above a given $\rho$ is, 
up to  a logarithmic factor which we neglect, $N(\rho)$.
 
Using the above estimate, the average energy and entropy of the thermal
atmosphere situated above a given $\rho$ are 
\begin{eqnarray}
\langle E\rangle &\simeq& N(\rho ) {1\over M} \simeq
{M\over \rho^2}\nonumber\\
S &\simeq& N(\rho)\simeq {M^2 \over \rho^2}
\label{EStotal}\end{eqnarray}
These qualitative result as well as the fluctuation of the 
energy estimated below have been obtained in quantitative detail
using the brick wall model, i.e., evaluating the partition
function of the Schwarzschild modes in the WKB approximation
\cite{th85}.

The entropy of the atmosphere diverges as $\rho \to 0$. This led
t'Hooft to suggest that there is a cutoff at $\rho =1$ so that the
entropy of the atmosphere would coincide with the Bekenstein-Hawking entropy.

The average energy $\langle E\rangle$ also diverges as $\rho$ tends
to zero.
This is intimately related to the renormalization of the energy
momentum in Schwarzschild background. Indeed one can show\cite{bd} that
the renormalized energy density, as seen in the frame of an infalling
observer, 
 is finite in Unruh vacuum.
This is because
the divergence of  $\langle
 E\rangle$ in Eq.(\ref{EStotal})
is  compensated by the negative and divergent  mean energy density in
 Boulware vacuum (the state containing no Schwarzschild particles).
Thus, after renormalisation, $\langle E^{ren}\rangle$
is finite and of order the Hawking flux $1/M$.
Note that the finiteness of $\langle E^{ren}\rangle$ depends on the
fact that each particle in the thermal bath is correlated to a partner
as in Eq.(\ref{Unruhvac}), i.e. it depends on the state of the field on
both side of the horizon.
We shall assume in this paper that as predicted by the semiclassical
theory,
$\langle E^{ren}\rangle$ is indeed finite.

On the other hand the thermal energy fluctuations of the atmosphere are not 
affected by the renormalisation since they are associated with the
decomposition of Unruh vacuum as a thermal density matrix of
Schwarzschild quanta and each term in this decomposition has physical
significance. 
These thermal fluctuations are proportional to the square root of
the number of particles in the thermal atmosphere
as for any thermodynamic system.
Therefore,  the
 uncertainty of the \w energy in the region between $\rho $ and
 $R\simeq 3M$ is
\be \de M\simeq \frac{1}{M}\sqrt{N(\rho)}\simeq\frac{1}{\rho}.\label{Uncert}\ee
Since the total energy of the black hole is fixed to be $M$, the
 uncertainty of the Schwarzschild
energy between $r=0$ and $\rho $ is also $ \frac{1}{\rho}$.
Note that this uncertainty is dynamical.
Indeed since $N(\rho )$ particles cross the surface $\rho $
in a time $\de t =M$, the time scale over which the mass
 fluctuate is also $\de t =M $.

This uncertainty is much larger than $\langle E^{ren} \rangle$ 
and is
much larger than the uncertainty due to the
emission of s-waves (see point 3 in the introduction). 
It implies the existence of strong gravitational interactions at $\rho
= \sqrt[3]{M}$. 
To see this let us first estimate
how the fluctuating mass gives rise to uncertainty in the location of the horizon.
A point $r_0$ is outside the horizon if 
\be \delta (r_0)=r_0-2M(r_0)>0,\label{out}\ee
 where $M(r_0)$ is the \w energy between $r=0$ and 
$r=r_0$. From 
Eq.(\ref{Uncert}) we find that 
\be \de \delta (r_0)=2 \Delta M
\simeq \frac{2}{\rho}=\frac{2}{\sqrt{8M\delta} }
.\label{out2}\ee
Clearly, if $\de\delta (r_0)> \delta (r_0)$ then the
 point $r_0$ is in a superposition of being inside  and outside
the horizon. From Eqs.(\ref{out2},\ref{rho},\ref{out}) this
 implies that the minimal $\rho $ for which it is
 certain that the point is outside the horizon is 
\be \rho_{min}\simeq\sqrt[3]{M}.\label{rhomin}\ee
The quantum fluctuations smear the horizon on an invariant
 distance of the order of $\sqrt[3]{M}$ which is much larger than $1$.

This suggests  that the gravitational
interactions with the atmosphere become large at $\rho =\sqrt[3]{M}$
and that the assumptions of a free field propagating on a given
classical background break down at this scale. We illustrate this
by inserting  the mass fluctuation $\de M$ into the \w metric. Near
the horizon one then obtains an
equation of the form
\be \p ^2_t-\left( 1-\frac{4\de M}{r-2M}\right) \p^2 _{r_*}+
\left( 1-\frac{2\de M}{r-2M}\right) \left(\frac{r-2M}{2M}\right)
\left( \frac{l(l+1)+1}{M^2}\right) =0.\label{kg1}
\ee
The perturbation is negligible as long as  
\be \frac{\de M}{r-2M}= {M \over \rho^3} > 1\ee
which yield $\rho >\sqrt[3]{M}$ as above.
For smaller $\rho $ the perturbation cannot be neglected.
Furthermore since $\de M$ varies over time scales of the order 
of the inverse particle energy, the solution of Eq.(\ref{kg1})
 will contain both positive and negative frequencies below 
$ \rho = \sqrt[3]{M}$ so the number operator $a^{\dag} a$
has uncertainty of order one.

The analysis in this section was based on the thermodynamics
of the fluctuating atmosphere. Gauss's law  then implied that
gravitational interactions occur on scales $\rho =\sqrt[3]{M}$.
The main drawback of these arguments is that they treat
$\de M$ as a classical source in Einstein equation rather 
than quantum fluctuation.
 However in the next section we shall recover this characteristic
 length $\rho = \sqrt[3]{M}$
 using a completely
different approach.

\section{The gravitational interactions}

In this section
we shall show that strong gravitational interaction occur at
$ \rho = \sqrt[3]{M}$ by studying the gravitational
interaction between an infalling particle and the thermal atmosphere.
Although the interaction between
each particle is small, there are approximately 
$N(\rho)$ such interactions which sum up incoherently.
Therefore the total probability of scattering is proportional
 to $\sqrt{N(\rho)}$. 
The final result is that at $\rho = \sqrt[3]{M}$ the total scattering 
probability is  of order 1.
However, due to technical difficulties, at the 
present time we have only investigated in detail the interaction
between an infalling s-wave and the high angular momentum particles of the
atmosphere. Similar results may hold for the interaction among the
high angular momentum particles themselves.
 
The detailed calculation is carried out in the appendix. Here we
summarize the results.
As shown in \cite{sex}\cite{th84}, the semi-classical gravitational
effects of a massless particle can be obtained using the
gravitational shock wave (the corresponding scattering amplitude 
coincides up to a phase with
one graviton exchange \cite{grav}).
We have considered the shock wave of an infalling particle
 with energy $E$  and its effect on the high angular momentum 
particles which constitute the atmosphere.
The probability that one particle of the atmosphere be in the same state 
after crossing the shock wave is
\be P_1\simeq 1-{\frac{M^2 \lambda^2}{\rho ^4}},\label{P1}\ee
where $\lambda $ is the energy of the infalling particle.
The number of   particles which are affected by the
shock wave of the ingoing particles when it reaches $\rho$
is given by $N(\rho)$, so the  probability for the 
``atmosphere'' above
$\rho $ to be in the same state is
\be P^{tot} =P_1^{N(\rho)}=(  1-{\frac{M^2 \lambda^2}{\rho ^4}})
^{N(\rho)}
\simeq  e^{-\frac{M^4 \lambda^2}{\rho ^6}},\label{Ptot}\ee
this means that for 
\be \rho <\lambda^{\frac{1}{3}} M^{\frac{2}{3}}\label{rhomin2}\ee
the probability for the   ``atmosphere'' to remain in the same
 state
decreases exponentially.

In the appendix we also show that the probability for the angular
momentum of one particle of the atmosphere not to have changed
coincides with $P_1$, i.e.,
\begin{equation}
P_{\Delta l =0} \simeq 1 - {\lambda^2 M^2 \over \rho^4}=P_1
\end {equation}
Hence proceeding as from Eq. (\ref{P1}) to
Eq. (\ref{rhomin2}), the angular momentum of the atmosphere is
modified by one unit when the particle reaches 
$\rho  \simeq \lambda^{\frac{1}{3}} M^{\frac{2}{3}}$.

Note that at $\rho =\lambda^{\frac{1}{3}} M^{\frac{2}{3}}$ (Eq.(\ref{rhomin2})),
the probability that any individual high angular momentum
particle be scattered is (see Eq.(\ref{P1}))
\be 1-P_1\simeq (\lambda /M)^{2/3}\ll 1.\ee
The weakness of the gravitational interaction justifies the
semi-classical treatment of the gravitational interaction.

The minimal $\lambda$ one can consider is $\frac{1}{M}$
since otherwise the wave length of the ingoing particle is
 larger than the radius of the black hole.
Therefore, at 
$\rho =\sqrt[3]{M}$
all   ingoing particles have interacted strongly with the
atmosphere  and have acquired one unit
 of angular momentum.
Although we have not been able to show it rigorously at this
 stage, we
expect that the ingoing particle will also be scattered by the
atmosphere (the principle of action and reaction) and 
that the high angular momentum particles which
 make up
the atmosphere will  be strongly self interacting at $\sqrt[3]{M}$.
This was indicated  by the analysis of the end of Sec.2
 wherein we naively plugged the fluctuating mass into the
 Klein-Gordon equation and estimated its effect on the 
propagation of a mode.

\section{The information problem}

In this section we consider the implications of our results for
the  S-matrix ansatz proposed by 't Hooft.
Since the interactions are strong at 
 $\rho=\sqrt[3]{M}$
this distance should  
play a 
crucial role in the  information problem.
Indeed, we shall show 
that the information of an 
ingoing
 massless charge-less spin-less
particle is encoded in the state of the atmosphere when the particle
reaches $\rho= \sqrt[3]{M}$\footnote{
We do not consider in this paper how the information about  the internal 
degrees of freedom of the particle, ie.
 its species, its spin state, etc.., are transmitted to the
 atmosphere.
The answer to this question is not clear at the moment since
 the information is transmitted via gravitational interaction 
(all other interactions are too small) and  the internal
 degrees of freedom couple weakly  to  gravity. }.

Such a particle is characterized by its energy and angular
momentum. We first consider how the information about its energy is
transmitted and then turn to the angular information.  
Consider an incoming particle in an s-wave $\Psi$ whose energy is $\lambda$ with
uncertainty $\de \lambda$.
This wave packet is spread out over an interval $\de t\simeq1/\de \lambda$.
 From the previous section we know that the incoming particle 
will start interacting with the particles in the atmosphere at 
$\rho =\lambda^{1/3} M^{2/3}$.
The particles in the atmosphere fall back towards the horizon after a
time interval $\de t\simeq M$.
Hence the interaction of the incoming particle with any individual
particle in the atmosphere lasts a time  $\de t\simeq M$.
One should therefore decompose $\Psi$ into a complete orthogonal set
of wave packets whose uncertainty in energy 
is $\de \lambda=1/M$ and that are spread out over a time 
$\de t\simeq M$.
We study how the information about the energy of two such wave packets
is encoded in
atmosphere.

Consider two particles whose wave packets have mean energy 
$\lambda_1$ and $\lambda_2$ and energy spread $\de \lambda_1\simeq
\de \lambda_2 \simeq1/M$. Since these particles are orthogonal we also have
$\lambda_2 - \lambda_1 > 1/M$.
We want to determined at what $\rho $ the state 
of the atmosphere when the energy of the
ingoing particle is $\lambda_1$ is orthogonal to
the state  of the atmosphere when the energy of 
the ingoing particle is $\lambda_2$.
To find this $\rho$ we 
 recall that the effect of the shock wave of an incoming
particle on a given particle in the atmosphere  is   a 
discontinuity in the $v$ direction. Using Eqs. (A4) and (A15), we get
\be \Delta v \simeq { \lambda M \over \rho} \log (\frac{\x}{M}),\label{Deltav}\ee 
 where  $ \x$ is the transverse distance between the ingoing
particle and a particle in the atmosphere (see the appendix).
Except when treating problems which explicitly involve the angular
momentum of the scattered particles, one can drop the logarithmic
dependence of Eq. (\ref{Deltav}).
Then the difference in the discontinuity is  only due to
  $\lambda_1\neq \lambda_2$: 
\be \Delta v_1 - \Delta v_2 \simeq 
{ (\lambda_1-\lambda_2) M \over \rho} . \ee From Eq.(\ref{P}) 
one obtains that the probability that the scattered state
of the particle in the atmosphere  is the same (i.e., does not
 depend on whether the energy of the ingoing particle is $\lambda_1$ or
 $\lambda_2$) is 
\be P_1\simeq1-\frac{M^2 \Delta \lambda^2}{\rho ^4},\ee 
where $\de \lambda=\lambda_1 -\lambda_2$.
The number of particles which are affected by the shock wave is
$N\simeq\frac{M^2}{\rho ^2}$ (Eq.(6)) so the probability that the scattered
state  of the whole atmosphere is the same is 
\be P=P_1^{N(\rho)}\simeq e^{-M^4 \de \lambda^2 \over \rho ^6}
\label{PorthE}\ee
Since $\de \lambda>1/M$ we obtain that when the ingoing particle
 crosses $\rho=\sqrt[3]{M}$
the probability that the  scattered
state  of the  atmosphere  is the same is exponentially small.
Therefore the information on the energy of the ingoing particle is
 encoded in the atmosphere at $\rho = \sqrt[3]{M}$.

Of course this is not  all the information since  there are 
orthogonal states of the ingoing particle with the same energy
but different angular location/momentum.
We will prove now that at $\rho=\sqrt[3]{M}$ the angular information
of the ingoing particle is also encoded in the atmosphere.

Consider a particle which falls radially into a black hole
 along the direction $\Omega_1$ with energy $\lambda$. 
Imagine now that the particle falls into the black hole with 
the same energy but along another
direction $\Omega_2$ with $\Omega_2$ sufficiently different from 
$\Omega_1$ so that the two initial states are orthogonal.
 This orthogonality condition implies that
 $\Delta \Omega > 1/ \lambda M,$
where  $\Delta \Omega$ is the
angle between the two directions $\Omega_1$ and $\Omega_2$. 
 We want to
know at what $\rho$ the state of the atmosphere when $\Omega =
\Omega_1$ is orthogonal to the state of the atmosphere when 
$\Omega =\Omega_2$, i.e., at what $\rho$ the information about
 the
angular direction $\Omega$ gets encoded in the state of the
 atmosphere.
Obviously, the most difficult case to distinguish is when
 $\Delta \Omega$ takes its minimal value 
\be \de \Omega =\frac{1}{\lambda M},\ee 
so we will consider that case.
Since the energy in both cases is the same, $\lambda$,
the difference in the shift of the $i$ particle in the 
atmosphere is

\begin{equation}
\Delta v_1 - \Delta v_2 \simeq { \lambda M \over \rho} \ln {\tilde x_{1i} \over
  \tilde x_{2i}}
\end{equation}
Where $\x _{1i}$ and $\x _{2i}$ is the transverse distance
 between the
infalling particle (at $\Omega_1$ and $\Omega_2$) and the
 particle $i$. From Eq.( \ref{P}) one obtains that the probability that the
scattered state of particle $i$ is the same in case 1 and 2 is
\begin{equation}
P_{ i} \simeq 1 - {M^2 \lambda^2 \over \rho^4} \ln  {\tilde x_{1i} \over
  \tilde x_{2i}}
\end{equation}
The probability that the scattered state of all the atmosphere 
 is the same  is
\begin{equation}
P = \prod_i P_{ i} 
\end{equation}
To evaluate the product we need to estimate $ \ln  {\tilde x_{1i} \over
  \tilde x_{2i}}$. We shall consider the case $\lambda \leq T_{loc} = 1/\rho$. 
There is a lower bound on $\x$ which is the size $1/\lambda$ of the wave packet
of the incoming particle and an upper bound which is the size
 of the horizon $M$.
 From Eqs.(22) it is clear that  ${\tilde x_{1i} \over
  \tilde x_{2i}}$ differs from 1 by an appreciable amount only if
$\tilde x_{1i} , \tilde x_{2i} < \Delta \Omega M$.
 Thus
\begin{equation}
P \simeq  \prod_{i\ such\ that\  \tilde x_{1i} , \tilde x_{2i} <
  \Delta \Omega M}
(1 - {M^2 \lambda^2 \over \rho^4})
\end{equation}
where we have neglected the log factor which is legitimate since
$\tilde x_i$ is bounded from below. 
Since the total
number of particles in the atmosphere  is $N(\rho)$, the number of
particles such that $\tilde x_{1i} , \tilde x_{2i} <
  \Delta \Omega M$ is $\Delta \Omega^2 N(\rho)$ and
one obtains
\begin{eqnarray}
P &\simeq&
(1 - {M^2 \lambda^2 \over \rho^4})^{\Delta \Omega ^2 N(\rho)
}\nonumber\\
&\simeq& e^{- M^2/ \rho^6}\label{PorthL}
\end{eqnarray}
 
Remarkably both Eq. (\ref{PorthE}) and (\ref{PorthL}) are 
independent of the initial energy $\lambda $ of the
infalling particle. Thus, provided that $\lambda \leq 1/\rho$, 
the information  about the energy and angular
position of an infalling particle
gets encoded in the state of the atmosphere at $\rho=\sqrt[3]{M}$.

\section{Conclusion}

The existence of a thermal atmosphere above a black hole is well
known. We have shown that this atmosphere plays an essential role in the gravitational
back reaction to Hawking radiation. Indeed it implies the existence of
strong gravitational interactions at $\rho =\sqrt[3]{M}$ and not at
$\rho =1$ as would be naively expected. This was shown by analyzing
the gravitational effects of the atmosphere.

In the first approach we estimated the thermal energy fluctuations of
the atmosphere. To estimate their effects, we then inserted these fluctuations 
as a classical source in 
Einstein's equations. This shows that the horizon seems to be
fluctuating on scales $\rho =\sqrt[3]{M}$.
We do not know if this effects would survive in a more careful treatment
of the gravitational interaction,
but in any case it indicates that  the propagation of
particles can no longer be described by a linear quantum field for 
$\rho <\sqrt[3]{M}$. In particular the decomposition of Unruh vacuum
as a thermal density matrix of non interacting particles is incorrect
for $\rho <\sqrt[3]{M}$.

In a second approach we calculated how the presence of an incoming
particle modifies the state of the atmosphere due to the gravitational
interaction. We find that the atmosphere gets scattered to an
orthogonal state before the particle reaches $\rho =\sqrt[3]{M}$. Once
more this shows that the atmosphere cannot be described as a gas of
noninteracting particles. 
However the full implications of this result cannot be
 understood at present
 because we have not been able to estimate how
the infalling particle is scattered by the atmosphere and how the
particles which constitute the atmosphere interact among themselves.

We then further investigated the interaction of an infalling
 particle
with the atmosphere and showed that  the information carried by
 the
infalling particle can get encoded in the atmosphere at $\rho
=\sqrt[3]{M}$. This confirms the critical role of
 $\rho =\sqrt[3]{M}$.

The main criticism that one can make at our approach is that 
we have
treated the constituents of the atmosphere as on shell particles
rather than vacuum fluctuations. Indeed we have first traced over the
partners before evaluating the gravitational response. This is however 
expected to be a valid
approximation if   the 
S-matrix ansatz of t'Hooft is correct, a fact which appears to be corroborated
by the analysis of section 4.
Thus our analysis implicitly implies a restriction to the region
outside the horizon. The question then arises of whether an
infalling observer can cross the horizon and fall into the singularity
as predicted by the semiclassical theory. In answering this question
the existence of partners beyond the horizon will play a crucial
role. Indeed it can be shown that in certain physical processes the 
presence of the
partners is essential in ensuring insensitivity to
the transplanckian frequencies which occur in Hawking
radiation\cite{MaPa}\footnote{
It is also interesting to note that since the analysis of the 
appendix was carried out in the
Rindler approximation, Rindler horizons defined over a transverse
distance $L \times L$ are probably 
 fluctuating on distances $\sqrt[3]{L}$. But in the Rindler case
 it is obviously possible to cross the horizon.}.

Nevertheless if one restricts oneself to the region outside the
horizon our analysis strongly suggests that there is a new phase at
$\rho =\sqrt[3]{M}$  where gravity becomes strongly coupled to the thermal
atmosphere.
Thus whereas  the Hawking radiation is ignited as in the
conventional free field theory, the source of the thermal radiation
progressively shifts  to   the  new phase   at
$\rho =\sqrt[3]{M}$.  This new phase can capture information of 
infalling matter and this
is in line with the idea that the black hole evaporation is 
unitary and that the
black hole entropy is stored in the thermal atmosphere outside the classical
horizon. 
It remains however to be seen whether the information about the star
that collapsed to form the black hole also gets encoded in the
atmosphere.  Possible
consistency could be achieved if the star itself does not 
collapse but becomes a
source of the burning atmosphere\cite{eng}.

Whatever the details of the physics near the horizon, the essential
result of our paper is that strong gravitational interactions already
occur at $\rho =\sqrt[3]{M}$ where the local temperature $T_{loc} =
1/\sqrt[3]{M}$ is  small. This may have important
implications for several proposed scenarios of black hole evaporation
which appeal to strong interactions at much smaller distances. Indeed
the S matrix proposed by t'Hooft 
neglects the high angular momentum particles and relies on
gravitational interactions which are strong only at the Planck scale\cite{th90}
and Susskind's picture of stringy horizons makes appeal to non
perturbative effects which should arise at the Hagedorn
temperature\cite{sus2}. It is still too early to understand 
the connection with the recent
advances in the string theoretic description of black
holes\cite{vafa,callan}. 

\vspace{1cm}

The authors would like to thank Y. Aharonov for his participation in a
fruitful discussion which was the original impetus for this
work. S. M. would like to thank T. Jacobson for discussions on related problems.

\begin{appendix}

\section{Appendix}

In this section we derive Eq.(\ref{P1}).
First let us briefly summarize the effect of the shock wave
(the full details are in \cite{th84}).
The gravitational field of a massless point like
particle in Minkowski space
 is described by  the line element \cite{sex}
\be ds^2 =-du(dv+2p^v\ln (\frac{\x^2}{M^2})\delta (u-u_0)du)+dx^2+dy^2,
\label{shock}\ee
where $\x ^2=x^2+y^2$, $u=T+z$ and $v=T-z$.
The massless particle moves in the $v$ direction with constant
 $u_0$ and momentum $p^v$ .

In Minkowski space there is an arbitrariness in the length
 scale appearing in the log which can be modified using the
 coordinate change $v\rightarrow v+\Theta (u-u_0)c$.
Since we are using Minkowski and Rindler coordinates to approximate
the \w metric near the horizon,
the curvature of the \w metric fixes the length scale in the log to be of order  
$M$. None of our results depend on the exact value of this length scale.

The effects of such a shock wave on other particles
are most easily analyzed in the action formalism.
The solution of the Hamilton Jacobi equation for a massless
 particle of initial momentum $k_{\mu}$ and
propagating in the metric (\ref{shock})
is
\be
S = S_0 + 2 p^v \ln (\frac{\x^2}{M^2}) k_v \theta(u-u_0) +
O((u-u_0)\theta(u-u_0) )\label{S}\ee
where $S_0$ is the solution in the absence of shock wave
\be
S_0 = k_x x + k_y y + k_v v + \frac{k_x^2 + k_y^2}{4 k_v} u
\label{S0}\ee
The effect of the shock wave is therefore a discontinuity in
 the $v$
direction at $u=u_0$:
\be v_0 = \frac{\partial S}{\partial k_v} =-\frac{k_x^2+k_y^2}{4 k_v^2}u+ v + 
 2 p^v \ln (\frac{\x^2}{M^2}) \theta(u-u_0) +
O((u-u_0)\theta(u-u_0))
\label{dis}\ee
and a refraction in the transverse direction:
\be
k_x (u) = \frac{\partial S}{\partial x} = k_x + 
 \frac{ 4 p^v}{\x^2}x k_v\theta(u-u_0) +
O((u-u_0)\theta(u-u_0) )
\label{ref}\ee
and similarly for $k_y(u)$.

One further verifies that the solution of the Klein Gordon
 equation in
the presence of the shock wave is given by 
$\psi = e^{iS}(1 + O((u-u_0)\theta(u-u_0))$. Thus the WKB
approximation correctly describes the effect of the shock wave.

Let us now use these results to describe how a high angular momentum
particle is affected by an incoming particle.
First let us recall that near the horizon and for transverse distances
smaller than $M$ the \w metric takes the approximate form
\be ds^2=-\frac{\rho ^2}{(4M)^2}dt^2+d\rho ^2+dx^2 +dy^2\ee
which is simply Minkowski space 
\be ds^2=-dudv+dx^2 +dy^2\ee
in Rindler coordinate
\begin{eqnarray}
\label{rindler}
u=T+z=\rho e^{t/4M}\\ \nonumber
v=T-z= - \rho e^{-t/4M}
\end{eqnarray}
\begin{figure}
\begin{picture}(400,400)(0,0)
\put(40,200){\vector(1,1){200}}
\put(40,200){\vector(1,-1){200}}
\put(100,260){\line(1,-1){200}}
\put(267,27){$u_0$}
\put(275,35){\vector(1,1){25}}
\put(265,25){\vector(-1,-1){25}}
\put(240,400){u}
\put(223,0){-v}
\put(260,110){ ingoing particle}
\put(180,240){particle in the atmosphere}      
\put(180,60){\line(0,1){120}}
\put(170,190){\line(0,1){140}}
\put(100,60){$\rho _0$}
\put(115,60){\vector(1,0){65}}
\put(95,60){\vector(-1,0){55}}
\end{picture}
\caption{The effect of the shock wave of the ingoing
  particle on Hawking particle with high angular momenta is a
  discontinuity in $v$.  The picture represents the trajectories
  projected onto the $u,v$ plane.}
\end{figure}
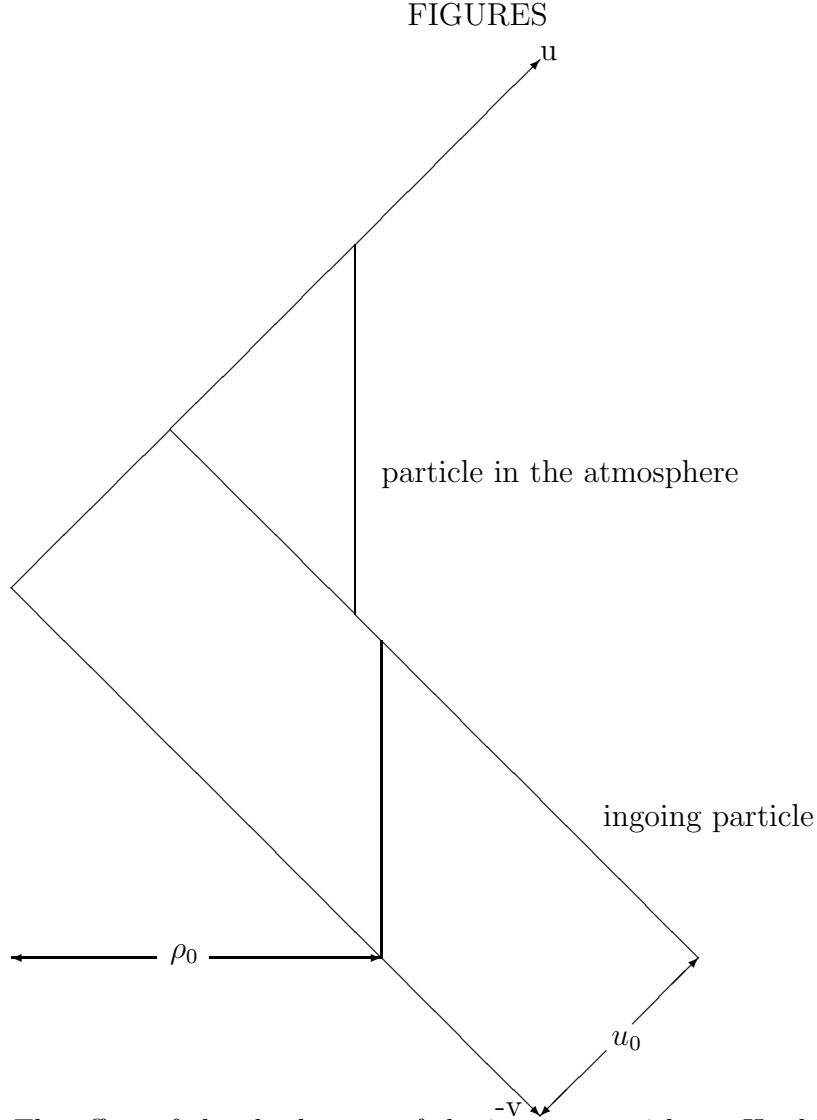
Particles in the atmosphere are massless and follow geodesics
\be X^{\mu}=X^{\mu}_0+\lambda V^{\mu}_0\ee
with $V_0^2=0$.
A boost in the $T, z$ plane corresponds to a translation in \w 
time.
By such a translation in $t$ and a rotation and translation
 in the $x, y$ plane we can bring the trajectory to the
 following form 
 (see Figure~1)
\be X^{\mu}(\lambda)=(T=-\rho _0 +\lambda , y=0,  
x=-\rho _0+\lambda , z=\rho _0)\;\;,\;\;0<\lambda<2 \rho_0,
\label{traj}\ee
 $\rho _0$ is the maximal $\rho $ the particle can reach,
so it is related to the angular momentum by
$\rho _0=\frac{M}{l}$ (see Eq. (\ref{maxl})).

The ingoing particle is moving along the line $u=u_0$.
It is easy to see that the high angular Hawking quanta will 
cross the shock wave of the ingoing particle at 
\be\rho _{c}^2=2\rho _{0}u-u^2.\label{rhocross}\ee
Most of the particles which reach the point $\rho_0$ will reach 
a maximal $\rho $ of the order of $\rho_0$.
Therefore we are interested in 
\be u\simeq \rho_0\simeq \rho_c. \label{36}\ee 

In order to be able to use Eqs.(\ref{dis}, \ref{ref}) 
we need to relate \w energies to 
Minkowski energies.
Denoting \w energy by $\lambda$ we obtain
\be \lambda= - p_{t}= - g_{t\mu}p^{\mu}=\frac{\rho ^2}{(4M)^2}p^t.\ee
Since $t=2M\ln (\frac{u}{v})$ we get
\be p_t= \frac{u}{4M}p_u-\frac{v}{4M}p_v, \label{pt}\ee
For the incoming particle $p_v =0$ so
\be - p_u=\frac{1}{2}p^v=\frac{4M\lambda}{u}\simeq \frac{M\lambda}{\rho
  _0}
\label{pv}\ee
where we have used Eq.(\ref{36}).
For the high angular momentum particle following the
 trajectory Eq.(\ref{traj}), $ k_v=k_u$, so Eq.(\ref{pt}) implies
\be - k_u = - k_v \simeq {\omega M \over \rho_0} \simeq
\frac{1}{\rho_0}\label{k}\ee where we have used the fact that the Schwarzschild
energy of particles in the atmosphere is $\omega \simeq 1/M$.

The wave packet which describe such a high angular momentum
Hawking particle is
\be \s{\phi}  =N\int dk f(k)e^{ik(x-T)}.\ee
Where $f(k)$ is a function such that $\de k=\overline{k}=
\frac{1}{\rho_0}$ and $N$ is a normalization factor.
 Eqs.(\ref{dis}) implies that after the wave packet crosses the shock
 wave there is a  discontinuity 
$\de T=p^v$.
Neglecting logarithmic factors we find  
that after crossing the shock wave the state of the Hawking
 particle is 
\be \s{\phi} ^{'} \simeq N\int dk f(k)e^{ik(x-T +p^v)}
\label{phi'}\ee
Therefore, the probability to be in the same state after
 crossing the shock wave is
\be P=\mid\langle\phi^{'}\s{\phi}\mid ^{2}\simeq 
N^2 \int dk \vert f(k)\vert^2 (1 - k^2 p^{v2})
= 1-p^{v2} \de k ^2\simeq
1-\frac{M^2 \lambda^2}{\rho_0^4}\label{P}
\ee
where we have used Eq. (\ref{pv}).

In addition to the shift in the longitudinal direction, the
angular momentum of the particles which constitute the atmosphere also
changes. 
In the classical trajectories this appears as the refraction
Eq. (\ref{ref})
\begin{equation}
\Delta p_x = {4 p^v k_v x \over \tilde x^2}
\simeq { p^v  k_v \over M}
\simeq { \lambda \over \rho_0^2}
\label{px}\end{equation}
where we replaced $\tilde x$ and $x $ by their typical value $M$ and
used the estimates of $p^v$ and $k_v$ obtained above.
The relation between $p_x$ and the angular momentum $l$ is $p_x = l/
M$,
hence Eq. (\ref{px}) corresponds to mean change of angular momentum
$\Delta l \simeq { \lambda M \over \rho^2} <<1$.
Because of the smallness of $\Delta l$, the corresponding change in
the wave function is
\be \s{l=l_0} \rightarrow \s{l=l_0} +i\frac{\lambda M}{\rho ^2} \s{l=l_0\pm
  1}
\label{deltal}\ee
Therefore the probability for one particle in the atmosphere to have
changed angular momentum is 
\be
P_{\Delta l \neq 0} = {\lambda^2 M^2 \over \rho_0^4}
\ee 
which coincides with
Eq. (\ref{P}).

Note that Eq. (\ref{deltal}) can also be obtained by noting that Eq
(\ref{phi'}) neglects the logarithmic dependence of $S$ and that the
scattered
modes are in fact
$e^{i k(x -T + {p^v} ln \tilde x^2)}$. The $\tilde x$
dependence of the log can be shown to imply Eq. (\ref{deltal}).

Thus we have shown how an s-wave interacts with the atmosphere.
At present  we cannot show how the high angular momentum
particles which constitute the atmosphere interact among each other.
The reason is that in order to do so one must know how to describe the
gravitational interaction between  two Hawking particles with high angular
momentum and how to describe the
gravitational effect of  Hawking particle with high angular
momentum on an outgoing particle.
But, unlike the ingoing particle, the high angular
momenta are just vacuum fluctuations in Minkowski space.
They correspond to  short lines (compared to $M$) 
in Minkowski space (see figure 1), and therefore their gravitational
effect cannot be approximated by the 
shock wave.

\end{appendix}

\end{document}